\documentclass[runningheads]{llncs}

\usepackage[T1]{fontenc}

\usepackage{graphicx}
\usepackage{cite}
\usepackage{amssymb,amsfonts}
\usepackage{algorithmic}
\usepackage{graphicx}
\usepackage{textcomp}
\usepackage{xcolor}
\usepackage{url}

\usepackage{enumitem}

\def\BibTeX{{\rm B\kern-.05em{\sc i\kern-.025em b}\kern-.08em
    T\kern-.1667em\lower.7ex\hbox{E}\kern-.125emX}}

\usepackage[framemethod=TikZ]{mdframed}
\newenvironment{intro}[3][]{
  \mdfsetup{
    skipabove=6pt,      % pads top.
    skipbelow=0pt,
    hidealllines=true, 
    leftline=true,      % left line
    linewidth=0pt,
    innerlinewidth=1pt, 
    innerlinecolor=#2,
    backgroundcolor=#3,
    innerleftmargin=10pt,   % Padding on the left
    innerrightmargin=10pt,  % Padding on the right
    innertopmargin=6pt,    % Padding at the top
    innerbottommargin=6pt,   % Padding at the bottom
  }
  \fontsize{10pt}{11pt}\selectfont % Adjusting font size within the frame
  \begin{mdframed}}
  {\end{mdframed}}

\newenvironment{zeroindent}
  {\par\setlength{\parindent}{0pt}}
  {\par}

\definecolor{red}{HTML}{e97132}
\definecolor{orange}{HTML}{fbe3d6}

\usepackage{array}
\newcolumntype{x}[1]{>{\centering\arraybackslash\hspace{0pt}}p{#1}}

\begin{document}

\title{Challenges of Virtual Validation and Verification for Automotive Functions}
\author{Beatriz Cabrero-Daniel\inst{1,3}\orcidID{0000-0001-5275-8372
} \and \\Mazen Mohamad\inst{2,3}\orcidID{0000-0002-3446-1265}}

\authorrunning{Cabrero-Daniel and Mohamad} %<3

\institute{University of Gothenburg \and RISE, Research Institutes of Sweden \and Chalmers University of Technology, Gothenburg, Sweden}

\maketitle

\begin{abstract}
Verification and validation of vehicles is a complex yet critical process, particularly for ensuring safety and coverage through simulations. However, achieving realistic and useful simulations comes with significant challenges.
To explore these challenges, we conducted a workshop with experts in the field, allowing them to brainstorm key obstacles. Following this, we distributed a survey to consolidate findings and gain further insights into potential solutions.
The experts identified 17 key challenges, along with proposed solutions, an assessment of whether they represent next steps for research, and the roadblocks to their implementation. While a lack of resources was not initially highlighted as a major challenge, utilizing more resources emerged as a critical necessity when experts discussed solutions.
Interestingly, we expected some of these challenges to have already been addressed or to have systematic solutions readily available, given the collective expertise in the field. Many of the identified problems already have known solutions, allowing us to shift focus towards unresolved challenges and share the next steps with the broader community.

\end{abstract}

\section{Introduction}

% ESTABLISHING THE CONTEXT, BACKGROUND AND/OR IMPORTANCE OF THE TOPIC
Verification and Validation (V\&V) of automotive functions is very challenging, mainly due to the huge amount of test scenarios to cover in order to ensure safety, which requires hundreds of years to perform in a physical setup \cite{kalra2016driving}. Hence, an established practice in automotive is to use simulations for safety and for ensuring coverage \cite{fadaie2019state}. 
However, achieving realistic and useful simulations is challenging. It is crucial to understand the limitations and obstacles in reaching this level of realism to effectively use simulations for V\&V \cite{sagmeister2024analyzing}.
%However, it is not easy to achieve realistic and useful simulations. It is super important to understand the challenges and limitations to achieve this level of realism so that we can use simulation for V\&V.

% GIVING A BRIEF REVIEW OF THE RELEVANT ACADEMIC LITERATURE AND IDENTIFYING A PROBLEM, CONTROVERSY OR A KNOWLEDGE GAP IN THE FIELD OF STUDY
%The white and grey literature have looked into existing approaches to using synthetic data for V\&V, however, there are no universally agreed upon standards for testing vehicles nor for developing and assessing the virtual toolchains.

Reported literature in the field (both white and grey) have explored various approaches to utilizing synthetic data for verification and validation (V\&V) in automotive testing \cite{cabrero2024digital}. However, a major issue remains: there are no universally accepted standards for testing vehicles or for developing and evaluating virtual toolchains. This lack of standardization leads to inconsistencies in how different organizations define, implement, and validate their simulation environments. Without common benchmarks and guidelines, it becomes difficult to ensure the reliability, accuracy, and comparability of simulation-based V\&V processes across the industry. 

% STATING THE AIM(S) OF THE RESEARCH AND THE RESEARCH QUESTIONS OR HYPOTHESES
%There was a project on these topics: a, b, and c. The aim of this study is to gather their experiences regarding the challenges and limitations to achieve this level of realism so that we can use simulation for V\&V. The study also asks experts for potential solutions to these challenges and evaluates their feasibility.
The aim of this study is to collect insights from experts regarding the challenges and potential solutions for simulation-based verification and validation gathered during a 3 year-long research project. These experts were both academics and practitioners from the whole automotive value chain, with a large expertise on the topic at hand, from eight different organizations, that worked on a project focusing on Enabling virtual validation \&  verification for automotive functions by utilizing physical test and simulations. Their suggested solutions are then assessed for feasibility, helping to identify practical approaches that could enhance the accuracy, reliability, and applicability of simulations in automotive testing and validation.
% PROVIDING A SYNOPSIS OF THE RESEARCH DESIGN AND METHOD(S)
%First we conducted a workshop with many brainstorming activities, then we sent out a survey to consolidate the results and gather further insights. 
The research process began with a workshop that incorporated various brainstorming activities to encourage open discussion and idea generation among participants. This initial phase allowed experts to share their experiences, identify key challenges, and group these challenges together into challenge groups. A survey was then sent out to further consolidate the results of the workshop and to gather additional insights and potential solutions to the identified challenges. 
This enabled us to address the following research questions:

\begin{enumerate}[leftmargin=0.9cm]
    \item[\textbf{RQ1}] What are the challenges of using simulations for validation and verification of driving functions? %Answered in Section~\ref{sec:challenges}. % answer: list of challenges from workshop, the grouping in table 1, challenge dedup
    \item [\textbf{RQ2}] What solutions exist to address these challenges and what roadblocks exist that hinder these solutions? %Answered in Section~\ref{sec:challenges}. % answer: ask them about simple problems, add: throw in more money!
    \item [\textbf{RQ3}] How common and generalizable are these challenges and how well known are the solutions? %Answer in Section~\ref{sec:rq3}, Table~\ref{tab:challenge_themes}, Figure~\ref{fig:incidenceofchallenges} and Figure~\ref{fig:nextfrontier}. % answer: votes from ws
   % \item RQ4: Which challenges have established solution and which constitute the next frontier for virtual V\&V? Answered in Figure~\ref{fig:nextfrontier}. % answer: ask them in survey?
    %\item RQ2.1: What roadblocks are there to achieve these solutions? Answered in Section~\ref{sec:challenges}. % answer: partly from discussions (tags in post-its)
   % \item RQ1.2: How can these challenges be categorized? Answered in Section~\ref{sec:challenges} % answer: the grouping in table 1
    %\item RQ1.3: To what extent do \textbf{these challenges affect the work on simulations} for AD verification (priority, criticality) % answer: should we ask them? how? via survey? simple-chaotic range, bea would drop this!
    %\item RQ1.3: Can these challenges \textbf{generalised to other projects}? % answer: ask different companies whether they encountered these in other projects
\end{enumerate}

% EXPLAINING THE SIGNIFICANCE OR VALUE OF THE STUDY
%The main take away of this study is the need to share the lessons learnt in these kind of projects so that we can stand on the shoulders of giants in the future and reach a higher technology readiness level for simulations for V\&V. Without this contribution, if everything stays in house, the know how does not pile up and the field does not advance.

The key takeaway from this study is the importance of sharing lessons learned from these types of projects. The industry can build upon past efforts and avoid repeatedly starting from scratch only by openly exchanging knowledge and experiences. This collaborative approach allows researchers and practitioners to stand on the shoulders of giants in the future and reach a higher technology readiness level for simulations for V\&V. Without this collective contribution, valuable insights remain within individual organizations, preventing the cumulative growth of expertise in the field.
% PROVIDING AN OVERVIEW OF THE DISSERTATION OR REPORT STRUCTURE

Section~\ref{sec:context} gives an overview of the context of this study (the project), Section~\ref{sec:RW} provides related work, and~\ref{sec:method} describes the research method for the study. Results are presented in Sections~\ref{sec:challenges} and~\ref{sec:rq3}, and the discussion is in \ref{sec:disc}. Finally, we conclude and discuss future work in Section~\ref{sec:conclusion}.

% Board 1: https://zoom.us/wb/doc/wcvK9yj4Tv6mTqfjZ9Vl9w/p/160628543455232 
% Board 2: https://zoom.us/wb/doc/ACczrJrHSn22029d-L34YA/p/160628543455232
% Board 3: https://zoom.us/wb/doc/2q37AGgSTEa9_rJvEt9pXg/p/159828994031616

%Challenge 1 - G13
%Challenge 2 - G2
%Challenge 3 - G6
%Challenge 4 - G14
%Challenge 5 - G11
%Challenge 6 - G1
%Challenge 7 - G8
%Challenge 8 - G5
%Challenge 9 - G15
%Challenge 10 - G4
%Challenge 11 - G9
%Challenge 12 - G16
%Challenge 13 - G3
%Challenge 14 - G7
%Challenge 15 - G10
%Challenge 16 - G10
%Challenge 17 - G12

%Survey questions mapping:
%This is a challenge that I have encountered (RQ1.2, RQ1.3)
%Can you tell us more about the challenge? (RQ1 in general)
%What was the degree of agreement among the participants regarding the challenge and the best way to address it? (RQ1.1)
%What solutions were proposed for addressing the challenge? (RQ2, potentially also RQ1.1)
%What do you think is the degree of certainty about what results will be generated from the solutions proposed for addressing the challenge? (RQ1.1)
%What roadblocks are there to achieve this solution? E.g., lack of documentations, not containerizing, etc. (RQ2.1)
%Where do we focus next? (RQ2.2) % 1.5 page
\section{Context}
\label{sec:context}

This study summarises the final activity of a research project, a workshop to reflect on the challenges faced during the project and gather lessons learned. This section briefly presents the project to serve as context for the present report. 
The purpose of the research project at focus was to explore and develop V\&V (validation and verification) strategies that balance feasibility and reliability of virtual test in a measurable way. The project considered both complementing and completely replacing traditional real-world data collection with simulated data. To effectively achieve this, the project conducted simulated and physical tests (in a test track) to quantify the gap between these two test environments. Being able to do so would enable systematic testing of software functions in vehicle systems before deployment.

\paragraph*{Participation} Nine partners, both industrial and academic, participated in the project. The partners included 
four automotive companies with expertise in simulation environments, two academic partners, in charge of equipping and managing the test vehicle, a research institute, a company managing a physical test track facility for AD and ADAS, who acted as coordinator, and a company specialised on gathering and annotating real data. The coordinator chaired weekly follow-up meetings in which progress was discussed, developing shared terminologies and understanding.

% \begin{figure}
%     \centering
%     \includegraphics[width=1\linewidth]{figures/evidentdiagram1.png}
%     \caption{Concepts addressed in the scope of the project.}
%     \label{fig:evidentdiagram1}
% \end{figure}

\paragraph*{Test vehicle} A Volvo XC90 with computers operating with the open-source OpenDLV software, which allows data collection from multiple sources (sensors) and control of brake, throttle and steering. The data gathered from the physical tests performed was shared with all project partners.

\paragraph*{Timeline} The first goal was to develop a common view on the best practices for gap analysis between simulated and real-world testing, as well as to understand each partners virtual toolchain and its maturity. Then, KPIs were discussed in a number of periodic workshops across the project. Five use cases were then distributed to the partners to perform physical and virtual tests using an open and transparent tool chain, which were used to compare the physical and virtual test and discuss the fidelity level achieved. In a final step, we organized a workshop (at the end of said project) to reflect on the challenges faced and lessons learnt, which are analysed in the present report, and to discuss their implications and potential solutions as of 2025. The workshop also helped the author structure the presentation of the findings.

\section{Related Work} \label{sec:RW}

% AUTOMOTIVE VERIFICATION AND VALIDATION
Testing is essential for the certification and safety of automotive functions. Therefore, it is essential to plan and design a V\&V strategy for each technology used in AD. Standards like ISO 26262 (about functional safety) provide strategies to mitigate systematic failures in hardware and software, as well as faults during design, implementation, verification, validation, and monitoring phases~\cite{ISO26262}. Artifacts that are used for conformance with these standards can then be used as evidence to argue for the functional safety for the function to V\&V. 

% SIMULATIONS TO DIGITAL TWINS
However, the infinite combinations of factors in their environment makes it impossible to cover them in physical test environments~\cite{18siddique2020safetyops}. Alternative strategies include testing in virtual environments: simulations are revolutionising testing in the automotive industry by enabling engineers to anticipate results, lowering costs and speeding up development~\cite{rasheed2020digital}. Digital Twins (DTs) expand on this by acting as virtual real-time mirrors of the systems and the test tracks~\cite{batty2018digital,cabrero2024digital}.

% SENSOR MODELS AND SYNTHETIC DATA
Simulations are based in models which provide mathematical representations of the systems under test~\cite{guala2002models}. The simulation models within the DT also mimic the behaviour of sensors such as cameras or LiDARs to capture the simulated environment~\cite{Conti:2009:DDS:1555009.1555162}, which should simulate realistic surroundings~\cite{AirSim,carla}, road networks~\cite{ISO23}, diverse weather conditions~\cite{nvidia2022digitaltwin,Ulbrich2015}, and dynamic obstacles~\cite{cabrero2024digital}.

% SCENARIO FILES AND OPEN FORMATS
Most DTs are typically built from scratch even though open-source software and tools exist to be reused~\cite{hu2023sim2real,cabrero2024digital} and to facilitate the integration of synthetic data into the V\&V process~\cite{cabrero2024digital}. By leveraging open-source tools, academic researchers can accelerate their research progress and contribute to the advancement of the field of DTs for V\&V~\cite{cabrero2024digital}.

% PHYSICAL TEST TRACKS AND THEIR DTS
Even though simulations and DTs are already being used to avoid risking lives and valuable equipment in early stages of V\&V~\cite{AIact,Ulbrich2015}, state-of-the-art virtual V\&V tools are not considered to be trustworthy enough on its own~\cite{unece}. This is not only due to technical challenges, but also organizational ones \cite{Mohamad2021}. Requirements Engineering (RE) therefore plays a vital role in ensuring that the DT accurately represents the real world (e.g., the physical test track facility) and can therefore provide useful insights into the vehicle system under test~\cite{cabrero2024digital}.

% FIDELITY GAP AND METRICS TO STUDY IT
Generating realistic data within a DT is as challenging as validating its accuracy~\cite{Conti:2009:DDS:1555009.1555162}. 
The use of real-world data enables the assessment of the fidelity of the DT in replicating the real world environment and the behaviour of the real vehicle system~\cite{Ulbrich2015}. By comparing the behaviour and performance of the DT with that of the actual vehicle, the virtual V\&V may be verified and validated~\cite{cabrero2024digital}. To achieve this, we need to establish appropriate standards to provide confidence in the reliability and safety of automotive features before testing in the real world~\cite{cabrero2024digital}.

% OPEN STANDARDS SUCH AS ASAM (V&V THE V&V TOOL)
Assurance cases have been used for a long time in various domains to reason about safety \cite{safetySAC} and cybersecurity \cite{SAC_SLR,cascade} and are explicitly required in various standards, e.g., ISO/SAE 21434 \cite{ISO21434} and ISO 26262 \cite{ISO26262}. 
The United Nations' ``New Assessment/Test Methods for Automated Driving (NATM)~\cite{natm} Guidelines for Validating Automated Driving System (ADS).'' NATM requires that ADS manufacturers provide evidence of the credibility of virtual toolchains~\cite{natm}. These documents would be used by relevant authorities to assess new automotive functions, formally V\&V them.

% ORGANISATIONAL CHALLENGES
%Despite the many technical challenges described in the literature, organisational challenges might be the most difficult ones to solve~\cite{MOHAMAD2024112082}. % 0.5 page
\section{Methodology} \label{sec:method}

To comprehensively identify and understand the challenges encountered in the research project, we organized a collaborative in-person workshop. This workshop brought together partners representing the full automotive value chain, including: Original Equipment Manufacturers (OEMs), providers, sensor suppliers, test site operators, and academics.
Following it, we conducted a survey\footnote{Details of the survey can be found in the Supplementary Materials: \url{https://doi.org/10.5281/zenodo.15798639}} to identify potential solutions, gather additional insights, and deepen our understanding of the identified challenges.
In the following sub-sections, we describe the activities conducted at the workshop and the survey and list the contributors in each activity.

\subsection{Collaborative Workshop}
The workshop brought together a diverse group of seven participants from five of the project partners (1 OEM, 3 test site operators, 1 academic, 1 sensor supplier, and 1 provider) as well as two moderators. We followed the \textit{1-2-4-All} Liberating Structure approach~\cite{liberatingstructures} to engage everyone in generating ideas and discussions. The approach starts with giving every participant a short time to reflect by themselves (1 minute), followed by a pair discussion (2 minutes) and lastly a group discussion (4 minutes). This is then done iteratively until a saturation is achieved. The total time of the workshop was 4 hours. As a tool for collaboration, we used the built-in collaboration features which allowed us to visualize the challenges as post it notes in a collaborative whiteboard.
The workshop consisted of three main activities:

\subsubsection*{Identification of challenges}
We started with asking each participant to individually list the challenges they had encountered during the project. This was followed by a pair and then a group discussion on these challenges. At the end of the activity, similar challenges were grouped together to streamline a list and identify patterns.

\subsubsection*{Voting on importance}
Once the challenges were organized, participants were asked to vote on the ones most important. Each participant was given 10 votes to distribute the challenges (max 1 vote per challenge). This activity helped highlighting the areas that required more focus or further discussion. On average, the participants used seven of their votes. As a result we got a list of ordered challenges based on their perceived importance for participants.

\subsubsection*{Categorization}
In the final step, we asked the participants to categorize the challenges based on two dimensions: the degree of agreement among the project partners regarding the way to address the challenge; and the degree of certainty on whether results will be generated from it. The goal of this activity was to let participants reflect on the solutions for the identified challenges and their complexity. This activity was done in two groups followed by a joint discussion.
As a result, challenges were categorized according to an Agreement-Certainty Matrix~\cite{liberatingstructures} into \emph{simple}, \emph{complex}, \emph{complicated}, and \emph{chaotic}. The Supplementary Materials include the categorization by each of the groups that was later used for discussion~\cite{supp}.

\subsection{Survey to all project partners}
To gain deeper insights into the collected challenges, we sent out a survey, published as Supplementary Materials \cite{supp}, that included Likert and open-ended questions about the nature of the challenges, their possible solutions, and potential roadblocks of the solutions.  
Additionally, we asked the respondents, listed in Table~\ref{tab:respondents}, to select the challenges that should be the focus in upcoming projects, and we also asked if they wanted to add additional challenges. 

\begin{table}[h]
    \centering
    \caption{Focus of the organisations of the survey participants}\label{tab:respondents}
    \vspace{-.1cm}
    \begin{tabular}{l|r}
        \textbf{Participant} & \textbf{Partner contribution} \\
        \hline
        P1, P7 & Academic \\
        P2 & Simulation, OEM \\
        P3, P4, P6 & Test site operator \\
        P5, P9 & Simulation \\
        P8 & Research institute \\
        P10 & Test vehicle provider \\
    \end{tabular}
    \vspace{-.5cm}
\end{table}

While the number of experts participants in the final workshop was relatively small, the authors would like to remind the reader that their insights reflect the work of the larger group, as well as the reality of the state of the art as of 2025.

%\begin{figure}
%    \centering
%    \includegraphics[width=0.5\linewidth]{figures/postitcluster.png}
%    \caption{Post-it cluster}
 %   \label{fig:postitcluster}
%\end{figure}

%Mazen's proposal for each challenge group:

%Challenge group GX, \{\{description\}\}, clusters some of the challenges identified during the workshop. 
%\begin{enumerate}
%    \item Did you encounter these challenges?
%    \item What is your point of view on this challenge group?
%    \item From 1 to 3, what is the degree of agreement among project partners regarding the way to address these challenges?
%    \item From 1 to 3, what is the degree of certainty and predictability about what results will be generated from the solutions proposed for addressing the challenge?
%    \item Any other comment?
%\end{enumerate}

%Bea's concern: how do we aggregate all these matrices? % 0.5 page
\section{Challenges and Proposed Solutions} \label{sec:challenges}

The problems that occurred during the project were identified by the workshop participants and written in sticky notes. During the second half of the workshop, the participants merged them to reveal 17 distinct challenges. This section presents each of them as described by the workshop participants, addressing \textbf{RQ1}, and classified into: technical, resource-related, missing requirements, and organizational.
Each is accompanied by the solutions proposed in the follow-up survey, addressing \textbf{RQ2}, as well as a discussion on the main roadblocks to achieve this solution, according to the experts. 

% \begin{table}
% \centering
% \caption{Challenges identified by the workshop participants (merging the sticky notes counted in the third column), and the experience with said challenges of the workshop participants.}
% \resizebox{\linewidth}{!}{
% \begin{tabular}{llx{0.9cm}x{1.1cm}x{0.9cm}}
% ID & Short challenge description & Merged notes & Experienced by & In other projects \\ \hline
% C1 & Ideal sensors are unrealistic & 4 & 100\% & 60\% \\ %\hline
% C2 & Distorsions in sensors & 1 & 50\% & 100\% \\ %\hline
% C3 & Malfunctions lead to missing data & 1 & 70\% & 86\% \\ %\hline
% C4 & Parsing complex data & 4 & 50\% & 60\%\\ %\hline
% C5 & Lack of synchrony in data sources & 1 & 90\% & 89\% \\ %\hline
% C6 & Measuring significance of results & 1 & 70\% & 57\%\\ %\hline
% C7 & Limited test track availability & 2 & 30\% & 100\% \\ %\hline
% C8 & Simulation realism gap & 1 & 70\% & 71\% \\ %\hline
% C9 & Represent full system behaviour & 4 & 70\% & 86\% \\ %\hline
% C10 & Realism needs can vary & 2 & 60\% & 50\% \\ %\hline
% C11 & Difficult to plan the use of sensors & 1 & 30\% & 67\% \\ % \hline
% C12 & Determining fidelity gap & 4 & 90\% & 44\% \\ %\hline
% C13 & Human drivers compared to AD & 1 & 50\% & 40\%  \\ %\hline
% C14 & Proprietary formats & 6 & 100\% & 50\%\\ %\hline
% C15 & Intellectual property concerns & 1 & 90\% & 44\% \\ %\hline
% C16 & Lack of sensor specifications & 1 & 60\% & 100\%\\ %\hline
% C17 & Changes in human resources & 1 & 70\% & 57\%\\ %\hline
% \end{tabular}%
% }
% \label{tab:challenge_themes}
% \end{table}

\subsection{Technical}
\vspace{-5px}
\subsubsection*{\textbf{\colorbox{orange}{C1}. Correlation between physical and simulated sensors}} is important in some levels of testing. However, ``inaccuracies have occurred in replicating a physical sensor'' (P9). Idealizing sensors removes an important aspect when verifying functionalities, as ``the behaviour of a system might end up being different'' (P2), e.g., real-life sensors can malfunction or perform poorly. Moreover, ``sensor models may not represent accurately the hardware's performance across weather conditions'' (P7).
\vspace{-5px}
    \paragraph*{Solutions}
    % not rely only on ideals sensors but having HIL, do full pipeline tests, do further research
    Conducting ``trial test runs to align simulation and measured sensor outputs'' could be a solution (P6). Another approach involves using more accurate sensor models or ``combining simulations with [hardware-in-the-loop strategies]'' (P8). A practical step is to perform a ``dry run to test out the whole pipeline'' from data collection to processing (P4). Key actions are to use ``concrete KPIs and standardization'' (P2), clarifying ``how detailed a model needs to be'' and setting ``limits on accuracy for specific tests'' (P5). 
 \vspace{-5px}   
    \paragraph*{Roadblocks}
    The ``lack of high-fidelity sensor models from suppliers'' (P6) makes it hard to adjust the sensor model to the needs of the tests as ``correct sensor models could only be developed by each sensor supplier'' (P8) and remain a ``black box'' for others (P2). This is mainly due to providers not revealing sensitive information, e.g., sensor limitations. 
    % Additionally, there is a ``lack of research on developing correct environments'' and understanding the ``computer capacity'' required (P8).
\vspace{-5px}    
\subsubsection*{\textbf{\colorbox{orange}{C2}. Unexplained distortions in sensors' log data}} happened during the project due to unpredictable factors such as ``drift, weather, or terrain'' (P6). This led to ``inconsistent logs between experiments and unexpected values which could be correlated to other collected data'' (P3).
\vspace{-5px}
    \paragraph*{Solutions}
    % repeat tests, perform (automated) checks to calibrate well
    Avoiding distortions in logged datacould be achieved by repeating the tests multiple times to have redundant data (P6, P10, P7) together with ``automated checks of the output logs to detect problems early could be some possibilities'' (P3). This could involve booking ``backup test track slots'' in case issues or anomalies are detected (P6). 
 \vspace{-5px}   
    \paragraph*{Roadblocks}
    %Hard to anticipate (occurs in post-processing only). Difficult to debug and reproduce
    P1 did however mention that ``sensor glitches may only occur in data post-processing, when [workarounds] are not possible anymore.'' Moreover, said anomalies could be caused by many factors, and it might be impossible to find a general solution.
\vspace{-5px}
\subsubsection*{\textbf{\colorbox{orange}{C3}. Sensor failures and malfunctions lead to missing data}} happened ``due to lack of knowledge about the sensors lent [led to having] a plan that could not be performed'' (P8), ``wrong measurements, and a loss of valuable time'' (P7). The workshop participants and survey respondents acknowledged how this became a challenge, given that the gathered data had ``missing data points in different test cases and a significant reduction of test cases where all results were available, making analysis difficult and results less certain'' (P3).
\vspace{-5px}
    \paragraph*{Solutions}
    % test beforehand, risk assessment
    Sensor failures or malfunctions could be mitigated, according to respondents, by preparing and testing the equipment better before running full-scale experiments (P3, P5) as well as taking these risks into consideration when planning the use of the test tracks (P8, P10), and ``if possible, using redundant systems'' to gather data (P3).
 \vspace{-5px}   
    \paragraph*{Roadblocks}
    % problems inevitable in real life testing, hard to keep track of many outputs, time (imposssible to redo)
    While this was seen as solvable to a large extent, ``malfunctions are inherent and inevitable in real-life testing,'' specially in early-development phases (P7, P10), and ``tight deadlines and too little time leads to mistakes; [moreover, we always try to get as much as possible] so they are always stressful'' (P3). This is mainly because it is often not possible to redo the experiments due to the limited resources (e.g., test track availability).
\vspace{-5px}
\subsubsection*{\textbf{\colorbox{orange}{C4}. Parsing complex simulation data for analysis}} required a lot of time ``due to different tool-chains [with] different outputs'' (P6).
The main challenge arises from processing simulated and real data measured in the test track to overlay them (P6, P9, P4, P10), since it might be otherwise difficult to compare ``since they do not share the same time steps'' (P4). However, using the terms of the workshop, P4 pointed out that this is a complex problem, rather than a complicated one, and P5 said it is ``always part of the job.'' 
\vspace{-5px}
    \paragraph*{Solutions}
    % common tools to parse and break down the data into simpler formats
    Common tools to parse and break down simulation and make it manageable were identified as the solution for fidelity and correlation analysis (P6, P8), and for model integration (P8). ``Regarding the timing of the data,'' P4 suggested to ``use the latest available data point'' instead of interpolating and overlaying the simulated and real data.
 \vspace{-5px}   
    \paragraph*{Roadblocks}
    % time, money, common documentation, access to data
    However, time, budget, and resources were identified as roadblocks (P6, P8, P10) to process the data appropriately, and better documentation was highlighted by P4 as needed ``to understand on a high level how data pre-processing might affect the result.''
\vspace{-5px}
\subsubsection*{\textbf{\colorbox{orange}{C5}. Consistent time-stamping}} for synchronizing simulated and real data was one of the roots of C4. Experts agreed on the need to time-stamp data (P6, P8) to make two time series (simulated and real, or from two sensors) comparable (P4, P5). The difficulty lies in finding a good starting point to do the synchronization (P10, P7), ``especially if [the signals] lead to the triggering of an actuator'' (P7).
\vspace{-5px}
    \paragraph*{Solutions}
    Consistent time-stamping would be simple if ``the logging was performed the exact same way in simulation as it is on the real vehicle'' (P4). %then why was there an issue?
    %Simple, just need to agree on a protocol (proper system design) or syncing in post-processing.
    However, some partners had to combine data ``in post processing'' (P5) by interpolating points, which was described as an ``engineering challenge that can be addressed by a proper system design'' (P1). This could include time-sync devices, using a network time protocol, or marking starts for scenarios (P2, P7, P10). To achieve this, P8 stated that ``more lopping and more testing'' would be needed. 
 \vspace{-5px}   
    \paragraph*{Roadblocks}
    %Resources and better documentation, interpolation might not be possible
    Unfortunately, in projects there might be ``sensor issues and a pressure to execute the measurements quickly'' (P6). Thus, more time, budget, and documentation would be needed (P8, P9, P4), as well as ``know-how regarding the setup of this kind of systems'' (P7).
\vspace{-5px}
\subsubsection*{\textbf{\colorbox{orange}{C6}. Test cases and simulation results must be significant and representative}} 
Sampling real-world scenarios and measuring what ``should be known'' is a major issue (P1): there is a tendency to simulate every possible combination of input variables, which can result in an overwhelming number of test cases (P6). Instead, we need to define ``what constitutes statistically significant'' (P8), but the limited number of runs per scenario ``makes it difficult to draw any statistically significant conclusions from the studies'' (P4).
 \vspace{-5px}
    \paragraph*{Solutions}
    % have more drivers and statistical measures in place
    Strategies to address this challenge include ``Defining criticality levels for test case simulations - this is usually connected to the ASIL levels'' (P6), ``ensuring that the simulation is within a certain statistical thresholds generated from the real world scenario runs.'' (P4), and ``making use of more drivers'' to perform physical tests (P7).
\vspace{-5px}    
    \paragraph*{Roadblocks}
    %Different expectations from different parties leading to diffrent scenarios considered important. More time and money!
    ``Environmental and physical factors are hard to fully simulate'' (P2).
    A major limitation is that ``it will take more time to run the tests, and will cost more to keep the [physical test] track'' (P4). Additionally, ``different stakeholders have different opinions on what scenarios are most valuable'' (P10).

%%%%%%%%%%%%%%%%%%%%%%%%%%%%%%%%%
%%%%%%%%%%%%%%%%%%%%%%%%%%%%%%%%%
%%%%%%%%%%%%%%%%%%%%%%%%%%%%%%%%%
\vspace{-5px}
\subsection{Lack of resources} % NOT A PROBLEM, BUT PART OF THE SOLUTION
\vspace{-5px}
\subsubsection*{\textbf{\colorbox{orange}{C7}. Availability of physical test track time}} was mentioned by all workshop participants. In the follow-up survey, P8 mentioned that the test track could be booked even before needs are elicited, and added that ``the challenge is even bigger for research projects'' due to funding.
Due to these difficulties, scenarios were changed and ``manual tweaks'' were needed to adjust the simulations multiple times (P6, P10), which lead ``to quick measurements with poor to no documentation'' (P6).
\vspace{-5px}
    \paragraph*{Solutions}
    %Better planning (time and resources)
    Better planning, including ``booking extra slots (buffer days) in case issues arise'' (P6, P8), was seen as the only solution for the scarce availability of the physical test tracks. 
    Moreover, in order to cope with the frequent changes in the studied scenarios due to using alternative tracks, P10 suggested working with auto-generated scenarios, and P6 highlighted the need to ``pre-align'' the simulation and physical test environment before the tests. 
 \vspace{-5px}   
    \paragraph*{Roadblocks}
    %additional work is needed to update the simulation env if (when) another road is used. Can be solved with better planning. 
    These solutions are expensive, particularly for research projects (P8). Moreover, no matter how much planning is done, ``there will always be [last minute] changes'' (P10) and ``sensor issues in the real [test] car'' (P6). These prevalent issues, that eat up the ``buffer time,'' could ``lead to less documentation due to lack of time'' (P6).
\vspace{-5px}
\subsubsection*{\textbf{\colorbox{orange}{C8}. The gap between models and real sensor data}}
% Simulation models lack correlation with real sensor data
``It is hard to get a sensor model that accurately models the real physical sensor and creates similar results'' (P9). Consequently, the used models are ``ideal,'' which means they do not model sensor distortions and never fail to detect obstacles or traffic agents. This is due to prevailing intellectual property (IP) issues, and was an issue both when testing each model and in full-system testing (P8).   
\vspace{-5px}
    \paragraph*{Solutions}
    % use high fidelity models (difficult), decomposing and iterating (however, not enough time or money)
    To narrow the gap, experts agreed that realistic simulation models would be needed. In order to achieve this, ``we need to decompose to lowest level of model'' (P8) and ``add different parameters until the results are good enough'' (P9), by comparing them to to real data (P8, P10). The same should be done ``to integrate several models'' (P8).
\vspace{-5px}    
    \paragraph*{Roadblocks}
    %time and money, access to high fidelity models
    As for other solutions, partners identified time, funding, and access to resources, in this case high fidelity models, as the main roadblocks to achieve this solution. Other factors include the difficulty to settle on ``KPIs to model and verify the level of realism'' (P1), and the difficulty to parametrize the environment (P10).

\begin{figure*}
    \centering
    \includegraphics[width=1\linewidth]{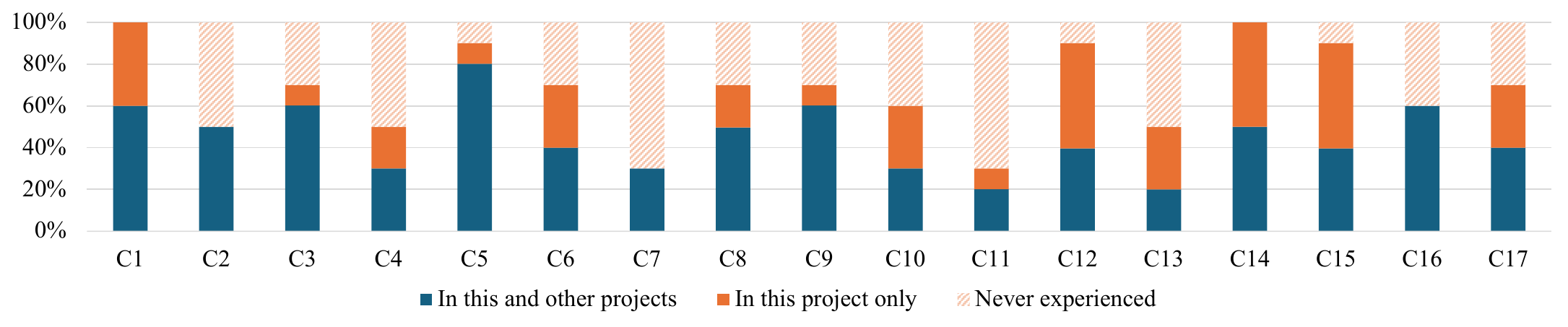}
    \vspace{-0.75cm}
    \caption{Proportion of participants having experienced each of the challenges in this project (orange) and either in this or in other projects (blue).}
    \label{fig:incidenceofchallenges}
\end{figure*}

\begin{figure*}
    \centering
    \includegraphics[width=1\linewidth]{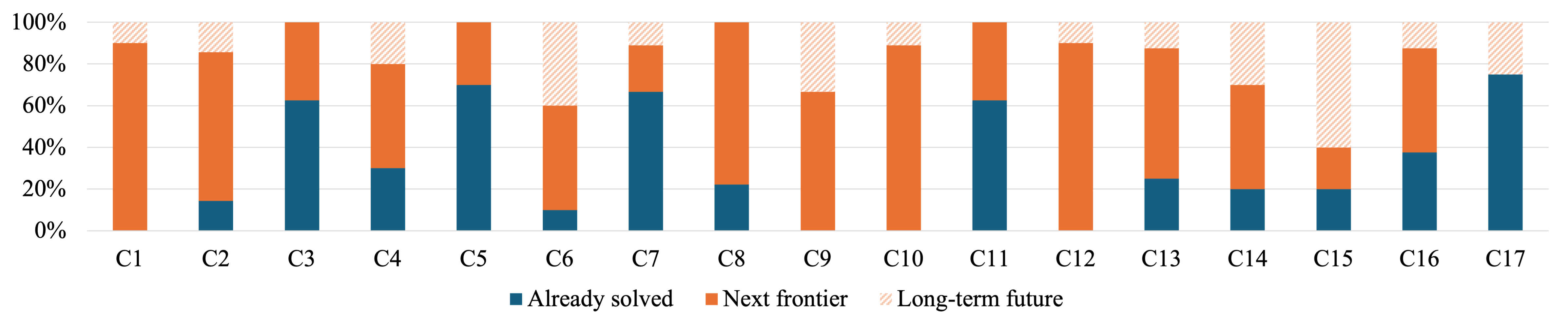}
    \vspace{-0.75cm}
    \caption{Classification of solutions (into solved, next frontier, or long-term future) for each identified challenge, according to the survey respondents.}
    \label{fig:nextfrontier}
\end{figure*}

%%%%%%%%%%%%%%%%%%%%%%
%%%%%%%%%%%%%%%%%%%%%%
%%%%%%%%%%%%%%%%%%%%%%
\vspace{-5px}
\subsection{Lack of requirements (missing or incomplete)}
\vspace{-5px}
\subsubsection*{\textbf{\colorbox{orange}{C9}. Accurate full system behaviour}}
This challenge is twofold. First, it is challenging to model complex vehicle systems with many components. Second, we need to assess the model's trustworthiness for debugging. Experts said that ``it is always a challenge to define the fidelity to opt for'' (P6) and to ``know when it is good enough, as there is no standard way'' (P4); it might depend on the vehicle (P10) and on the ``status of the project'' (P6): ``at lower levels of testing, [there is no issue] because this will always be handled higher up in the [V\&V] cycle and will be an internal OEM or TIER 1 discussion'' (P8), while higher levels need external assessors.
\vspace{-5px}
    \paragraph*{Solutions}
    % define test method (external) that includes some KPIs and acceptance criterias for the full systemi
    The respondents suggested external tests and assessment methods, in which clear expected outcomes, comprehensible acceptance criteria, and specific KPIs for realism are pre-determined (P1, P2, P6, P8).
    P6 suggested using ``different levels of fidelity depending on requirements and feasibility,'' and P4 said ``we should aim to reduce the complexity'' and ``make it easier to set thresholds for the [individual components'] simulation models.''
\vspace{-5px}
    \paragraph*{Roadblocks}
    % very complex system with many parts which might not be accessible and the environment is complex too (e.g., physical principles in evnrionment), all of these different areas might affect the results in the full system, a lot of dependencies
    This challenge was identified as unsolved by the experts, as seen in Figure~\ref{fig:nextfrontier}.
    The main reason is that there is a ``huge amount of models that need to be validated before the [full system] can be assessed'' (P8) and that it is difficult to determine how each affect the overall results; for instance, P2 mentioned ``understanding and accurately modelling all physical principles affecting sensor readings''. It is worth noting that the experts also discussed, in the workshop and in the survey, that there is typically a lack of access to and knowledge about the full vehicle system.
\vspace{-5px}
\subsubsection*{\textbf{\colorbox{orange}{C10}. Fidelity needs depend on scope and phase}}
The experts discussed at length the needs for simulation fidelity depending on project scope and phase within V model. P6 said that ``due to different status of [virtual] and [physical testing] one should not expect the best correlation from the start.'' As the project advanced, different tests were planned conducted: from ``simple unit testing or high level safety testing'' to ``the highest level of testing, [which] requires a higher fidelity model'' (P4). However, it is difficult to assess which fidelity level is acceptable when the requirements are not explicitly defined, as discussed during the workshop.
% This challenge was encountered due to the lack of requirements for the required fidelity levels during the definition of the project. This led to different interpretations among partners and hence different expectations of the required fidelity level. 
% In general, the challenge highlights the importance of specifying simulation fidelity expectations and requirements early in the project to ensure consistency and alignment among partners.
\vspace{-5px}
    \paragraph*{Solutions}
    % better planning (clear needs and scope for partners) and documentation, studying which fidelity level is needed for each V-phases (we need traces between requirements, related to the scope, and tests)
    This challenge can be tackled with better planning and documentation, particularly regarding the needs and scope of work for each partner. Additionally, each phase of the v-model can have  different a fidelity level and that needs to be traced back to the scope, requirements and tests of that particular phase. In this regard, P4 said that ``a new research project to investigate the mapping from testing in the V[-model] to the required fidelity.''
    %To the question What solutions were proposed for addressing the challenge? about challenge C10, P6 said "\textbf{Different model fidelity levels} and expected correlation levels across different V cycle stages". P8 said "To \textbf{define the needs of each partner} and agree how to handle the issue.". P4 said "A new research project to investigate the mapping from testing in the V to the required fidelity and to \textbf{reduce the complexity} in this question.". P10 said "\textbf{Better planning}. \textbf{consensus on project scope} and development phase.". P4 said "Documentation on what different levels of testing in the V should be applicable for the study. It should also be clear if it is applicable to AV or ordinary manual vehicles. ". 
\vspace{-5px}    
    \paragraph*{Roadblocks}
    Misunderstandings about the scope between partners led to this, which, according to P6, is ``always a challenge'' since requirements might evolve, and the physical vehicle and its digital twin might be at different stages of development and testing. Moreover, different partners might have different focus and goals, and hence different expectations of the needed fidelity.  

\vspace{-5px}
\subsubsection*{\textbf{\colorbox{orange}{C11}. Planning the use of sensors}} was difficult due to the lack of clear data requirements. As P3 pointed out in the survey: ``requirements on data are difficult to define, and might not be available in time when data collection needs to be done, which means some guesswork will be involved.''
\vspace{-5px}
    \paragraph*{Solutions}
    % having clear and evolving requirements
    Clear data requirements to plan the use of sensors, and for what to record or measure would be needed (P6, P10). P8 suggested using ``a checklist [while in the test track] and reviewing data live.'' P5 also pointed at routines and checklists, but stressed there are not ``fail-safe.'' 
 \vspace{-5px}   
    \paragraph*{Roadblocks}
    % evolving requirements 
    Once more, the respondents identified timing as a challenge to accomplish these solutions; for instance, P3 stated that ``it is difficult to fully specify data requirements before you have attempted to analyse the data, but you need data to start analysing.'' Because of this, an iterative approach with evolving requirements was recommended. 
\vspace{-5px}
\subsubsection*{\textbf{\colorbox{orange}{C12}. Determining the fidelity gap with meaningful metrics}}
is challenging, yet essential to achieve the goal of using simulations and DTs for V\&V. P6 said we must ``define what metrics will be reviewed and what is a good enough outcome.'' Therefore, there is a need quantify the gap and to reach a consensus for acceptable results among all partners. 
\vspace{-5px}
    \paragraph*{Solutions}
    % break down system into components, focus on parts, decide abstraction level
    As a solution to this challenge, the participants suggested breaking down the system into smaller and more manageable components, which allows for an easier decision regarding the required fidelity gaps and metrics. For instance, P2 said we need to ``break down a system into its components, following further down, and associating sensor outputs with metrics in combination with object properties.'' Finally, P5 said ``it would be good to find some limits on fidelity for what is needed for certain tests.'' 
\vspace{-5px}
    \paragraph*{Roadblocks}
    Lack of documentation and existing standards being too abstract and high-level. Additionally, there is a lack of established methods for credibility assessment of virtual tool chains along with KPIs to make sure that they are credible enough to perform V\&V. For example, P3 said ``there are proposed frameworks, e.g., in UNECE proposed NATM [but] leaves the question open of how to derive meaningful metrics and required fidelity thresholds for specific use cases.''

\vspace{-5px}
% simulating human driver or not?
\subsubsection*{\textbf{\colorbox{orange}{C13}. Whether to simulate a human driver or an AD}} lead to a discussion on the challenges of modelling human behaviour. ``There's a difference between the performance of an AD and a human,'' P7 wrote, as performance depends on the drivers, which ``will lead to a [higher] variability.'' Partners had different views on whether it was necessary to simulate human driving, and P5 wrote: simulating ``AD and human drivers answers different questions,'' so ``both are important.'' 
\vspace{-5px}
    \paragraph*{Solutions}
    % Unclear requirements can be solved with better planning and definition of scope.
    Defining whether to simulate an AV or a human driver early on and planning accordingly was defined as key both during the workshop and in the follow-up survey. P4 said the ``specific goals for the AV part of the study and for the human driver part of the study'' needed to be made clear, and P7 connected setting such goals with the scope of the verification, which also needed to be defined earlier on.
 \vspace{-5px}   
    \paragraph*{Roadblocks}
    %no roadblocks but it is difficult to model human drivers!
    Workshop participants also discussed the difficulty of simulating human drivers, which relies on ``including more subjective opinions and experiences'' (P5).

%%%%%%%%%%%%%%%%%%%%%%%%%%%
%%%%%%%%%%%%%%%%%%%%%%%%%%%
%%%%%%%%%%%%%%%%%%%%%%%%%%%
\vspace{-5px}
\subsection{Organizational}
\vspace{-5px}
\subsubsection*{\textbf{\colorbox{orange}{C14}. Integrating proprietary formats}}, still requires significant effort, despite industry standards. ``Simulation tool chains use different formats'' (P8) which need to be verified, but ``documentation about data is sometimes missing or outdated'' (P1), which is tedious. While standards exist, workshop participants said it is difficult to ``manually tweak'' them to fit the real test track (P9, P10). ``Scenario files [sometimes] need pre-processing work [to be imported] in commercial software'' (P6, P4), which ``always requires work than expected'' (P5, P10), and ``the existing set of standards and simulators is ever expanding,'' so ``it will be impossible to unify them'' (P4).
\vspace{-5px}
    \paragraph*{Solutions}
    % unify standards (ISO, NCAP), create interfaces and adapters and mappings, agree on best practices
    % who to decide and how to agree? there will always be new tools and formats and standards (e.g., for different granularity levels or for different parts such as the environment)
    Integrating data formats and platforms could benefit, according to the workshop and survey participants, from relying more on standards like ISO 26262 or the FMI standard, test catalogs like Euro NCAP, and agreeing on best practices (P6, P10, P2). 
    % SCENARIOS
    P3 explained a solution is ``creating a common platform for scenarios'' even though there already exists ``a number of existing tools using different open and [proprietary] file formats.''
    % INTERFACES
    Similarly, standard interfaces, APIs, and adapters should be provided (P8, P3); even though P3 points out that ``a 1:1 mapping is often not possible.''
\vspace{-5px}    
    \paragraph*{Roadblocks}
    % ROADBLOCKS
    It is difficult to establish ``the granularity in which the standards must be applied to the simulation,'' (P2) e.g., ``how to deal with surrounding environment, trees, buildings and so on?'' (P5), and NCAP scenarios were deemed too simple for their blank surroundings.
    %AGREEING
    P6 and P8 questioned who should lead the standardisation efforts, and pointed out the considerable effort needed to ``establish one [format] that works consistently for all commercial software.'' 
\vspace{-5px}
\subsubsection*{\textbf{\colorbox{orange}{C15}. IP concerns}} are a recurrent challenge when sharing ``sensitive highly detailed models [and] parametrizations'' (P5) between organizations because ``it could be possible to gain insight into how a AD stack works from them'' (P4). P6 said: ``IP frameworks vary across companies and resolving such discrepancies can take months or years.'' As a result, vehicle, sensor, and tire models were unlikely to be shared (P10, P7) or shared as a black-boxes to protect trade secrets (P3).
\vspace{-5px}
    \paragraph*{Solutions}
    %simplification and abstraction to overcome sensitive data sharing, share black-box models 
    ``Simplifying the whole simulation model, or parts of it'' (P4, P2), sharing only black boxes that can ``used without compromising IP'' (P7), or ``normalizing results so that the absolute values of the models is never presented'' (P4) were some of the suggestions. However, as seen in Figure~\ref{fig:nextfrontier}, solutions are considered long-term.
\vspace{-5px}    
    \paragraph*{Roadblocks}
    %No standard for how to abstract. Sensitive data is impossible to share. agreeing is hard, it is a case-to-case thing. It is hard to make sure that an abstracted model is good enough.
    The lack of agreement, e.g., on the level of abstraction to use (P2, P7) and how ``good enough'' should the shared models be (P5), was raised once more as a roadblock to achieving the proposed solutions. 
    Thus, P3 said that these tasks will always be decided on a ``case-to-case basis [and that] the issue largely remains for every new exchange.'' Moreover, P6 pointed out the ``long time [needed] for legal alignment'' and P3 added that ``new legal frameworks continually makes sharing more difficult.''
\vspace{-5px}
\subsubsection*{\textbf{\colorbox{orange}{C16}. Lack of detailed sensor specifications}}
IP concerns make it hard to get detailed sensor specifications, since only black box models are shared (P10) which makes it difficult to develop accurate simulations. Without knowing what is the output of physical sensors, it is difficult to ``align the simulation accordingly'' (P6) and simply defining ``low, mid and high'' fidelity levels is not enough (P8).
% could this problem also happen between teams in the same organisation?
\vspace{-5px}
    \paragraph*{Solutions}
    % sharing
    Respondents P6 and P8 wrote that there should be a higher alignment between partners, including sharing models. P8 added that ``sensor developers [must be] ready to take responsibility and clearly define fidelity gaps,'' but P3 thinks verification tasks need to be ``redefined as there is no good way to obtain a good enough model.'' 
\vspace{-5px}
    \paragraph*{Roadblocks}
    % company secrets and intellectual property
    The workshop and survey participants acknowledged that IP restrictions, and the protection of trade secrets will always be a challenge: sensor suppliers will still ``keep detailed performance characterization a secret'' (P3).
\vspace{-5px}
\subsubsection*{\textbf{\colorbox{orange}{C17}. Changes in human resources}} 
The experts reflected on how [human] resources changed out during the project. When this happens ``there is a loss of information'' (P4) and delays in the project, since there is not enough hand-over and ``ramp up time'' (P6, P3) and some tasks might be started from scratch, ``sometimes several times'' (P3).
% P8 said: ``As normal several persons have been changed out of different reasons during the project but in this project it have take to long for each partner to man up and come up to speed..'' 
% P3 said: ``Yes this has happened during Evident and other projects. Typically there is not enough hand-over and extensive learning time which means the tasks almost need to restart, sometimes several times during a project..''  P7 said: ``Loss of momentum and know-how.''
\vspace{-5px}
    \paragraph*{Solutions}
    % proper project management, redundancy, and documentation
    Better sub-project management, work documentation, task handover, and planning were the solutions proposed by the survey participants. This, the respondents wrote, should be the responsibility of each partner. Moreover, the ``project coordination could request a redundancy plan from each partner,'' P8 and P3 suggested.
\vspace{-5px}
    \paragraph*{Roadblocks}
    % hard to maintain and access documentation (specialized knowledge)
    All the survey participants agreed that changes in human resources were to be expected; P10 even said ``I see this as a risk in all projects.'' Moreover, they stated that documentation is hard to maintain and access, ``due to specialized knowledge needed, and the limited available funded time'' (P3). These issues were thus identified as roadblocks for these theoretical solutions in practice.
    % To the question What roadblocks are there to achieve this solution? about challenge C17 (Lot of [human] resources have been changed ot dur- ing the project.), P8 said "Life". P4 said \textbf{"(it is hard) Making it easy to read and change this documentation}. ". P10 said "I see this as a \textbf{risk in all projects}. Roadblock could be documentation in case of hand overs.". P3 said "In practice the solutions have not worked that well. People will move. The projects would need to have more persons involved to increase redundancy but that requires more funding.".In practice this is \textbf{difficult due to specialized knowledge needed} and limited available funded time and limited availability of persons.".  

%%%%%%%%%%%%%%%%%%%%%%%%%%%%%%%%%%%%
%%%%%%%%%%%%%%%%%%%%%%%%%%%%%%%%%%%%
%%%%%%%%%%%%%%%%%%%%%%%%%%%%%%%%%%%%
\vspace{-5px}
\section{General Challenges and Known solutions} \label{sec:rq3}
\vspace{-5px}
As workshop and survey participants pointed out, many of the challenges listed in Section~\ref{sec:challenges} can be seen as a ``risk in all projects.'' This section moves on to discuss which of these challenges, are common in other projects, and how well-known are their potential solutions, addressing \textbf{RQ3}.

As seen in Figure~\ref{fig:incidenceofchallenges}, all survey participants have experienced challenges C1 and C14, often in other projects too. Other challenges are almost as common: C5, C12, and C15 were also common, with 90\% of the survey participants having experienced it. Challenges C3, C6, C8, C9, and C17 were also common across partners and projects, as seen in Figure~\ref{fig:incidenceofchallenges}.
Some others were not as common, as they are typically experienced by specific partners, e.g., C7 and C11, which are very related to the use of the physical test track.

Figure~\ref{fig:nextfrontier}, gathers the opinion of experts on what is the next frontier for enabling virtual V\&V of automotive functions. A take away from this analysis is that a number of the challenges are, according to experts, already solved. Meanwhile, other challenges are left for the long-term future. It is also worth noting that for some of the challenges, participants did not agree on whether the challenges were solvable.

% 2 pages
% \section{Example challenge}

% \subsection{Challenge name}

% Challenge description.

% Challenge solution. \textit{Roadblocks.}

\section{Discussion}
\label{sec:disc}
% challenges identified (RQ1)
% In this project we have identified 17 challenges for virtual verification and validation for automotive functions through a workshop with the participants of a research project. The 17 challenges can be grouped in 4 categories: technical, organizational, managerial, and requirements-related issues. Some of these issues have a lot in common, for instance, the "it works on my machine" syndrome and the problems with working with physical test vehicles, that the experts claim are unavoidable.
This study has identified 17 challenges for virtual V\&V of automotive functions by asking representatives of the 10 partners of a research project, described in Section~\ref{sec:context}. These challenges were categorised into four groups: technical, organisational, managerial, and requirements-related. Several of these challenges, moreover, share common characteristics. For instance, the difficulties related with working with test vehicles, that the experts deem inevitable, as well as the ``it works on my machine syndrome.''

% while technical challenges, most were not
% While there were technical challenges, those were only 30\% of the total number of challenges; most of the these were related to organizational, managerial, requirements issues. For instance, a lot of things were unclear which led to many challenges, and which could have been mitigated by agreeing on following some standards and working towards measuring the verification of requirements using agreed upon KPIs.
Technical challenges accounted for only 30\% of the total identified challenges. The majority were related to organization, management, and requirements. For example, a common source of challenges was a lack of clarity that, according to experts, is unavoidable in early-stage development, but which could have been mitigated by establishing standard practices and verification requirements. 
% let us have evolving requiremets to address as things come up
% Even though eliciting measurable requirements and KPIs in the beginning is very difficult, a potential solution would be to periodically and systematically revisit and iterate over the requirements to avoid ending up with conflicting requirements. Agile method for research projects? Mari did very well, however, more resources might be needed to make things better.
Although doing so in early stages is universally challenging, a potential solution discussed was to periodically and systematically review and refine the requirements to prevent inconsistencies and conflicts as the project advances. Implementing a formal agile approach in research projects could facilitate them and support project management in guaranteeing the expected outcomes.

%we didn't lack resources, but having more is better
% Even though the lack of resources was not a major source of challenges, when the experts started proposing solutions to these challenges, utilizing more resources came up as a significant necessity. For instance, they mentioned more time, more funding, more human resources, more redundancy, more standards, etc.
\begin{intro}{red}{red!20}
The lack of resources was not identified as a major challenge, but it became evident when proposing solutions that they are a critical requirement for the continuation of this study, e.g., increased time, funding, personnel, redundancy, and standardised practices to adhere to.
\end{intro}

% we would expect some of these challenges to have never been there given collective know how (RQ2)
% We would expect some of these challenges to have never been there or have systematic solutions in store given the collective know how: we know what the solutions for many of these problems are. Challenge C6, related to ensuring that test cases and simulation results are statistically significant and representative of real-world scenarios, was described as long-term by the participants. This is surprising given the large body of work in the academic literature covering this challenges. Then, why did we face these challenges when we have encountered them in previous projects?
Given the collective expertise in the field and the large body of work in the academic literature covering these challenges, one might expect not to have encountered these challenges in the project. This work however reflects the reality of the state of practice as of 2025 and aims to remind the reader about the necessity to focus on the particular areas here discussed. The striking result, rather than the concrete challenges, is that these challenges still exist. This is specially true given that the systematic solutions to these issues seemed to be well-understood and agreed among partners. However, many of these, as seen in Figures~\ref{fig:incidenceofchallenges} and~\ref{fig:nextfrontier}, were identified as well-known yet long-term future challenges.

\begin{intro}{red}{red!20}
This raises two important questions: first, why do these challenges persist despite previous experiences? Second, if a similar workshop had been conducted before the project start, would the partners have reached an agreement on the solutions to implement?
\end{intro}

% even with challenges, evident went great, would it have been greater by analysing these first (based on experience from previous projects)?
% Even though we faced all these challenges, this project delivered great results for each of the work-packages and will serve as a basis for a subsequent project focusing on A and B.
% Very uncommon to see such an activity in a project plan, but perhaps it should be.
% Share challenges and knowledge. Good to have a lessons learned activity after the project, but A question then remains: would these challenges still be challenges if we had conducted a similar workshop before the project started? Probably not.
% please, keep sharing your experiences so we can learn from one another and move on to long term things (RQ2)
Despite the challenges encountered, the project successfully delivered valuable outcomes across all work packages, and will serve as a basis for a subsequent project focusing on the assessment of the virtual toolchains used for V\&V of automotive functions. Based on the learnings in this project, we will encourage incorporating a structured challenge analysis in the initial planning activities. Similarly, we also encourage researchers and practitioners undertaking similar projects to share their challenges and insights, fostering a culture in which we can collectively advance knowledge.

% now we have directions for future work (RQ2)
% The analysis, reported in this paper, of the encountered challenges and potential solutions, greatly inspired the lines for future research. Do not be ashamed to do the same in your projects and share your issues with the world so we can learn from them and not repeat your mistakes. Now, after getting over the already solved challenges, we can focus on the identified next steps and share them with the world.
% Moreover, this project has been crucial to shape future research directions, based on the solutions identified as next frontier by experts, presented in Figure~\ref{fig:nextfrontier}. We also strongly encourage others to share their 

\subsection{Threats to the validity of the results}

%In terms of Construct Validity, we consider the risk of misunderstandings and misinterpretations by the participants of the workshop and survey. Hence, we presented the purpose of the study and provided a brief context at the beginning of the workshop. In the second part of the workshop, they went through the challenges and disused it. Moreover, this is based on a project with partners that have worked together for at least two years, and developed their shared terminologies and understanding.
%For internal validity, we consider the workshop. We had two moderators that made sure that every participant got to share their opinions without any domination of an individual or a sub-group. We also used the 1-2-4 technique to give the participants time to reflect on their own before getting into the discussions (eliminating bias). For the survey, we did a filtering of the quotes to balance the contribution of respondents.

%For external validity, we asked the participants if they had experienced the challenges before, but we only got their own opinions. Hence, there is a risk that the findings cannot be generalizable. However, we had experts from different organizations both in academia and industry and the participants had many years of expertise in the field and good knowledge of the existing body of knowledge.

In terms of construct validity, we considered the risk of misunderstandings and misinterpretations by the participants in the workshop and survey. To mitigate this, we presented the purpose of the study and its context at the beginning of the workshop. In the second part of the workshop, participants discussed the identified challenges in detail. Moreover, this study is based on partners that have worked together for at least two years, developing shared terminologies and a common understanding of the subject matter.

For internal validity, we ensured a structured discussion during the workshop. Two moderators facilitated the session, ensuring that all participants shared their opinions without any dominating the conversation. Moreover, we employed the 1-2-4 technique~\cite{liberatingstructures}, allowing participants to reflect individually before the group discussions, reducing bias. For the survey, we carefully filtered the quotes to balance contributions, ensuring that no perspective was overrepresented.

Regarding external validity, we asked participants whether they had previously encountered the identified challenges, but we could only rely on their subjective opinions. As a result, there is a risk that the findings may not be fully generalizable. However, our study involved experts from diverse organizations, spanning both academia and industry. The participants also had many years of expertise in the field and a strong understanding of the existing body of knowledge, which strengthens the credibility of our findings.

%Technical: 1, 6, 10,13, 14, 17 

%Solved: 7, 10, 11, 14, 15, 16 % 0,5 page
\section{Conclusion and Future Work}
\label{sec:conclusion}

% RESTATING THE AIMS OF THE STUDY
%There was a project on these topics: a, b, and c. The aim of this study is to gather their experiences regarding the challenges and limitations to achieve this level of realism so that we can use simulation for V\&V. The study also asks experts for potential solutions to these challenges and evaluates their feasibility.
The aim of this study was to collect insights from experts regarding the challenges and potential solutions for simulation-based verification and validation of automotive functions. These experts were both practitioners and academics from the organizations that worked on a project focusing on enabling virtual V\&V for automotive functions by utilizing physical test and simulations.
% SUMMARISING MAIN RESEARCH FINDINGS AND SUGGESTING IMPLICATIONS FOR THE FIELD OF KNOWLEDGE
By doing this, we identified challenges, potential solutions, and roadblocks to achieve them. Moreover, we report expert opinions to understand how generalisable these are and whether they should be the next steps in research.

%Their suggested solutions are then assessed for feasibility, helping to identify practical approaches that could enhance the accuracy, reliability, and applicability of simulations in automotive testing and validation.

%This study aimed at identifying challenges and potential solutions for simulation-based V\&V. It is based on a project that focused on  

Future projects (academic and industrial) should build on these by providing systematic approaches to solve the ``already solved'' challenges, and focus on next frontier solutions. Specially, our experts stressed the need to establish a common framework for performing credibility assessments of virtual tool-chains and environments, which could promote virtual V\&V to complement physical testing. Additionally, this will support the use of simulations in regulatory testing, type approval, and certifications such as EuroNCAP.
%This is both for practitioners working on simulation for V\&V testing  of AD/ADAS, and also for researchers to have an outlook of the future research areas. 

% RECOGNISING THE LIMITATIONS OF THE CURRENT STUDY WHILST STATING A FINDING OR CONTRIBUTION

% MAKING RECOMMENDATIONS FOR FURTHER RESEARCH WORK AND SETTING OUT RECOMMENDATIONS FOR PRACTICE OR POLICY
%Even though this was outside of the scope of the project, experts agreed on the need to establish a common framework for performing credibility assessments of simulation chains and environments will support the use of simulations in regulatory testing, type approval, and certifications such as EuroNCAP. But will also give the trust for simulations needed to kick off the broad usage and possibility to replace physical testing. %from evident report % 1 page including references

\section*{Acknowledgment}
This work is supported by Sweden's innovation agency, Vinnova, under Grant No. 2021-05043 entitled ``Enabling Virtual Validation and Verification for ADAS and AD Features (EVIDENT)''.

\bibliographystyle{splncs04}
\bibliography{main.bib}





% LNCS Homepage, \url{http://www.springer.com/lncs}, last accessed 2023/10/25
% \end{thebibliography}
\end{document}